\documentclass[aip,jcp,reprint,letter]{revtex4-1}

\usepackage[pdftex]{graphicx}
\usepackage{framed}
\usepackage{amsmath}
\usepackage{amssymb}
\usepackage[all,cmtip]{xy}
\usepackage{amscd}
\usepackage[utf8]{inputenc}
\usepackage{pdfsync}

\newcommand{\tlabel}[1]{\label{t:#1}}
\newcommand{\fig}[1]{Fig.~\ref{f:#1}}

\newcommand{\sect}[1]{Section~\ref{s:#1}}
\newcommand{\tab}[1]{Table~\ref{t:#1}}
\renewcommand\vec[1]{\boldsymbol{#1}}

\newcommand{\rvec}{\vec{r}}
\newcommand{\Deltavec}{\vec{\Delta}}
\newcommand\diff[1]{{\rm d}#1}

\newcommand{\elabel}[1]{\label{e:#1}}
\newcommand{\slabel}[1]{\label{s:#1}}
\newcommand{\eq}[1]{Eq.~(\ref{e:#1})}

\newcommand\binaryLiftingVariable{h}
\newcommand\xunitvec{\vec{e}_{\mathrm x}}
\newcommand\yunitvec{\vec{e}_{\mathrm y}}
\newcommand{\tsim}{T_{\text{sim}}}

\newcommand{\cfigure}[3][\columnwidth]{
   \begin{figure}[!htb]
   \begin{center}
      \includegraphics[width=#1]{#2}
      \caption{#3}
      \label{f:#2}
   \end{center}
   \end{figure}}
\newcommand{\maxzero}[1]{ {[#1]}^+}

\newcommand{\flow}[3][]{\varphi^{\text{#1}}_{#2 \to #3}}
\newcommand{\TO}[2]{#1\! \to \! #2}

\begin{document}

\date{\today}

\title{Generalized event-chain Monte Carlo: Constructing rejection-free global-balance algorithms from infinitesimal steps}

\author{Manon Michel}
\email{manon.michel@ens.fr}
\author{Sebastian C. Kapfer}
\email{sebastian.kapfer@ens.fr}
\author{Werner Krauth}
\email{werner.krauth@ens.fr}
\affiliation{Laboratoire de Physique Statistique, Ecole Normale Sup\'{e}rieure,
UPMC, Universit\'{e} Paris Diderot, CNRS, 24 rue Lhomond, 75005 Paris, France}

\begin{abstract}
  In this article, we present an event-driven algorithm that
  generalizes the recent hard-sphere event-chain Monte Carlo method
  without introducing discretizations in time or in space.  A
  factorization of the Metropolis filter and the concept of
  infinitesimal Monte Carlo moves are used to design a rejection-free
  Markov-chain Monte Carlo algorithm for particle systems with
  arbitrary pairwise interactions.  The algorithm breaks detailed
  balance, but satisfies maximal global balance and performs better
  than the classic, local Metropolis algorithm in large systems.  The
  new algorithm generates a continuum of samples of the stationary
  probability density. This allows us to compute the pressure and
  stress tensor as a byproduct of the simulation without any
  additional computations.

\end{abstract}


\maketitle 

\section{Introduction} 

Markov-chain Monte Carlo (MCMC) methods in statistical physics have
progressed far from the original local-move, detailed-balance
Metropolis algorithm\cite{Metropolis_1953}. On the one hand, intricate
non-local cluster moves have met with great success in lattice models
\cite{Swendsen,Wolff}.  To a lesser extent, continuum systems of hard
spheres have in recent years also benefitted from non-local moves
\cite{Bernard_2009,Bernard_2011,Engel_2013,Kapfer_2013}, building on
earlier work\cite{Dress,Luijten,Jaster}. On the other hand, extensions
of the classic detailed balance condition have allowed to construct
Markov chains that converge faster.  These algorithms introduce
persistence between subsequent moves and reduce the diffusive nature
of the Markov chain on small and intermediate length and time
scales. Notable examples are guided random walks
\cite{Gustafson_1998}, hybrid Monte Carlo
\cite{Duane_1987,Peters_2012} and overrelaxation \cite{Neal_1995}. The
Markov chain lifting framework
\cite{Diaconis_2000,Andersen_2007,Chen_1999} unifies these concepts by
augmenting the physical configuration space with auxiliary variables
that resemble the momentum in Newtonian time evolution and in
molecular dynamics (MD)\cite{AlderWainwrightMD}.  Lifted Markov chains
have already been applied to spin models \cite{Turitsyn,Weigel}, but
not to continuum systems.

The present article draws on the above lines of research.  As a main
theoretical result, we introduce a factorized version of the
Metropolis filter (acceptance rule) that is well suited for the
simulation of $N$-particle systems with pair-potential
interactions. Combined with the concept of infinitesimal Monte Carlo
moves, this filter allows us to construct a rejection-free event-chain
Monte Carlo (ECMC) algorithm that breaks detailed balance yet
satisfies maximal global balance.  This algorithm builds upon a recent
insightful hybrid Monte Carlo scheme\cite{Peters_2012}.  By virtue of
infinitesimal displacements of particles, our algorithm produces a
continuum of configurations that all sample the equilibrium
distribution.  Samples are obtained efficiently using an event-driven
algorithm.  For hard spheres, the events correspond to hard-sphere
collisions, and the new algorithm reduces to the hard-sphere
event-chain algorithm\cite{Bernard_2009}.  For general pair
interactions, our algorithm replaces the hard-sphere collisions by
pairwise collisions, whose collision distance is resampled after each
event from to the pair potential.  Finally, the continuum of ECMC
samples permits to directly compute the pressure and the stress tensor
in the $NVT$ ensemble at no extra computational cost.

\section{Balance conditions, factorized Metropolis filter} 

\cfigure[85mm]{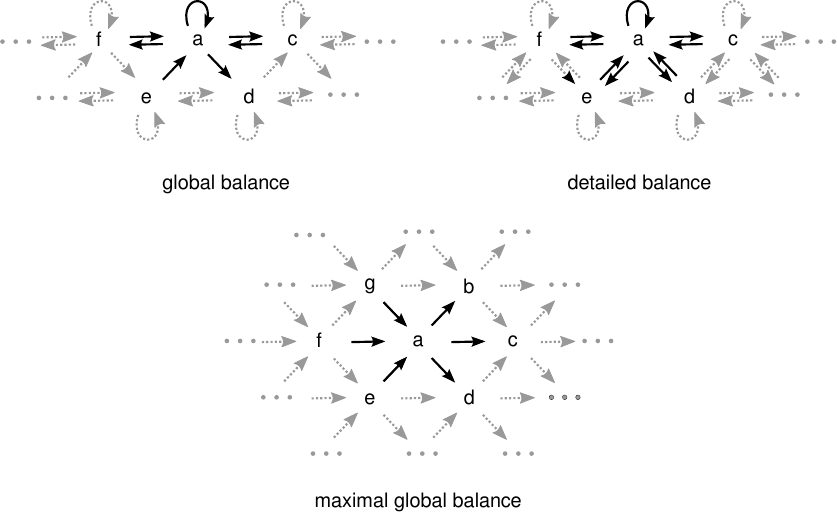}{Balance conditions for probability
  flow in Markov-chain Monte Carlo.  Arrows represent stationary flows of equal magnitude.
  \emph{Top left}: Global balance, a necessary condition for the convergence
  towards equilibrium. The total flow $\sum_c \flow{c}{a}$ into any
  configuration $a$ must equal the total flow $\sum_c \flow{a}{c}$ out of it.
  The loops $\flow{a}{a}$, etc., correspond to rejected moves.
  \emph{Top right}: Detailed balance: the net flow between any two
  configurations is zero, $\flow{a}{b} = \flow{b}{a}$.
  \emph{Bottom}: Maximal global balance: $\flow{a}{b} > 0$ implies
  $\flow{b}{a} = 0 $, the flow $\flow{a}{a}$ vanishes.}
  
For an MCMC algorithm to converge to the stationary distribution,
it must satisfy the global balance condition for the stationary flows
$\flow ab$ from configuration $a$ to $b$: the total flow
into a configuration $a$ must equal the total flow out of it,
\begin{equation}
   \sum_{b} \flow{b}{a} = \sum_{c} \flow{a}{c} = \pi(a),
\elabel{GlobalBalance}
\end{equation}
where $\pi(a)$ is the statistical weight of configuration $a$, e.\,g.~given
by a Boltzmann factor.
The flow must also satisfy an ergodicity requirement\cite{SMAC}.
The global balance, \eq{GlobalBalance}, is in particular satisfied by
the \emph{detailed balance} condition which equates the flows between any two
configurations $a$ and $b$: 
\begin{equation}
   \flow{a}{b} = \flow{b}{a}
\end{equation}
(see \fig{balance_conditions}).
We will be concerned with algorithms satisfying \emph{maximal global
balance}, where flow between two configurations is unidirectional
and flows from $a$ to $a$ (that is: rejections) are avoided: if $\flow{a}{b} > 0$, then
$\flow{b}{a} = 0$.  In this case,
probability does flow back nonlocally from $b$ to $a$. In the particle systems that
we consider, this happens \emph{via} the periodic boundary conditions.

MCMC methods commonly rely on the Metropolis algorithm, which enforces 
detailed balance of the flows between  $a$ and $b$ as follows:
\begin{equation}
   \flow{a}{b} = \mathcal A_{a \rightleftharpoons  b} \min(\pi(a),\pi(b)),  
\elabel{eqphi}
\end{equation}
In our algorithm, the a-priori probability $\mathcal A$ is symmetric
and amounts to zero or a global constant that we drop for simplicity.
\eq{eqphi} is manifestly symmetric in $\pi(a)$ and $\pi(b)$ so
that, by construction, $\flow{a}{b} = \flow{b}{a}$.  Since $\flow{a}{b} = \pi(a) 
p(a\to b)$, with $p$ the acceptance probability, \eq{eqphi} is equivalent to the well-known Metropolis filter
\begin{align}
   p(a \to b) = \min\left(1, \frac{\pi(b)}{\pi(a)} \right)
\elabel{eqprob}
\end{align}
that has been implemented in countless computer programs.

In statistical physics, the weight of a configuration $a$ is often given by
the Boltzmann factor $\pi(a) =\exp( - \beta E(a))$, 
where $E(a)$ is the energy of $a$ and $\beta$ is the inverse
temperature, which we set to one for the majority of this article.
Using the abbreviation
\begin{equation}
\maxzero{x} : = \max(0,x)\quad (\ge 0),
\elabel{minzero_definition}    
\end{equation}
we can write the Metropolis filter of \eq{eqprob} as 
\begin{equation}
    p(a \to b) =  \min(1,\exp(- \Delta E)) = \exp\left(-\maxzero{\Delta E}
\right),
\end{equation}
where $\Delta E = E(b) - E(a)$. 
This corresponds to the acceptance probability of a proposed move, whereas the
rejection probability is 
$    1 - p = 1 - \exp\left(-\maxzero{\Delta E}\right)$.

We now consider an $N$-particle system with pair interactions $E = \sum_{\{i,j\}}
E_{ij}$, where $i$ and $j$, in our applications, label particles in
$D$-dimensional space, but could also refer to spins or other degrees of
freedom. The sum runs over all unordered pairs $\{i,j\}$ of particles.
For such a system, the Metropolis filter has always been used as
\begin{align}
p^{\text{Met}}(a \to b)
  &=  \min(1, \exp (- \sum_{\{i,j\}}\Delta E_{ij}))\notag\\
  &=  \exp \Bigl(-\maxzero{\sum\limits_{\{i,j\}}\Delta E_{ij}} \Bigr).
  \elabel{normal_metropolis}
\end{align}
In the present article, however, we introduce a \emph{factorized Metropolis filter}
\begin{align}
    p^{\text{fact}}(a \to b) & = \prod_{\{i,j\}} \min(1, \exp (- \Delta E_{ij}))\notag\\
       & = \exp \Bigl( -\sum_{\{i,j\}} \maxzero{\Delta E_{ij}} \Bigr)
    \elabel{pair_metropolis_filter}        
\end{align} 
which also fulfills detailed balance by respecting the same flow
symmetries as the standard Metropolis filter, as can be seen by
applying the identity $\pi(a)/\pi(b) = \exp
(\sum_{\{i,j\}}\maxzero{\Delta E_{ij})}-\maxzero{-\Delta E_{ij}})$ to
\eq{pair_metropolis_filter}.  The conventional and the factorized
Metropolis filter agree in the hard-sphere case\cite{footnote_2} and
(trivially) for $N=2$ .  They differ whenever terms of opposite sign
appear on the rhs of \eq{pair_metropolis_filter}, i.\,e.~for general
interactions and $N>2$.  The factorization increases the rate of
rejections in a detailed-balance MC algorithm.  We find that for the
soft-sphere interactions considered in this article, the rate of
rejections is about 50\% higher (soft spheres with $n=12$,
$\rho=0.8\dotsc 1.2$, with a step size of $0.1$ in units of the
particle diameter).  However, the factorization yields the acceptance
probability as a product of independent pair interaction terms.  This
will be the key to derive a rejection-free lifted MCMC algorithm for
general $N$-particle systems.

\section{Lifting: 1D systems and two particles in a box}
\slabel{SingleParticle}

We now introduce the concept of Markov chain lifting in a simple
setting, which we later generalize to interacting particle systems.
We consider a one-dimensional discrete system with configurations $a$ and stationary weights $\pi(a)$ (i.\,e., $a \in [\Delta,2 \Delta, \dots,L\Delta]$). For moves sampled uniformly from $\{-\Delta, \Delta\}$,
the standard Metropolis
filter of \eq{eqprob} satisfies detailed balance $\flow{a}{a+\Delta} =
\flow{a+\Delta}{a}\ \forall a$. The stationary distribution $\pi$ is sampled in
the limit of infinite running times.

Lifting\cite{Diaconis_2000,Andersen_2007,Chen_1999}, in this example, consists
in duplicating each configuration $a$ with a momentum-like variable into two
configurations $a_\pm = (a,\binaryLiftingVariable = \pm 1)$.  The lifting variable
determines the next proposed move, which would in ordinary Metropolis MC
be sampled from a prior distribution: For $a_+$, only the particle
move $\TO{a}{a + \Delta}$ is proposed, and for $a_-$, only  $\TO{a}{a -
\Delta}$. For flow balance, we introduce
\emph{lifting moves} $a_+ \to a_-$ and $a_- \to a_+$ which take effect if
the particle move is rejected, as summarized in the diagram, 
\begin{equation}
\xymatrix{
\cdots \ar[r] &  (a-\Delta)_+ \ar[d]  \ar[r]^{\hspace{15pt}\varphi_0} &
a_+ \ar[r]^{\hspace{-15pt}\varphi_1} \ar[d]^{\maxzero{\varphi_0 - \varphi_1}} &
(a+\Delta)_+
\ar[d] \ar[r] &  \cdots\\
\cdots & \ar[l] (a-\Delta)_- \ar@<+4pt>[u] & \ar[l]^{ \hspace{15pt}  \varphi_0   }
\ar@<+4pt>[u]^{ \maxzero{\varphi_1 - \varphi_0} } a_- & \ar[l]^{\hspace{-15pt}
\varphi_1 }
\ar@<+4pt>[u] 
(a+\Delta)_- & \ar[l] \cdots \\
}
\elabel{lifting_diagram}
\end{equation}
where the flows $\varphi_0$ and $\varphi_1$ are given by
\begin{align}
\varphi_0 &= \min[\pi(a), \pi(a-\Delta)],\\
\varphi_1 &= \min[\pi(a), \pi(a+ \Delta)].
\end{align}
We take the weights of the lifted configurations to be the same
as the weights of the original configurations, $\pi(a_\pm)=\pi(a)$, adjusting
the constant of normalization.
This rejection-free MCMC algorithm satisfies maximal global balance, as
only one of the two flows $\flow{a_+}{a_-}$ and $\flow{a_-}{a_+}$ can be non-zero.
In the physical variables $a$, however, rejections are still present.

We now consider uniform stationary probabilities ($\pi(a) = \text{const}$)
and impose hard-wall boundary conditions in our one-dimensional discrete model.
The lifting flows are non-zero only for
a rightmoving particle at $a = L \Delta$, and for a leftmoving particle at
$a = \Delta$. For these configurations, the lifting flow equals the entire incoming
flow, and the particle reverses its direction. One can show that the lifted algorithm visits all sites
in $\mathcal O(L)$ steps, rather than in $\mathcal O(L^2)$ steps for the Metropolis
algorithm.\cite{footnote_3}

\cfigure[85mm]{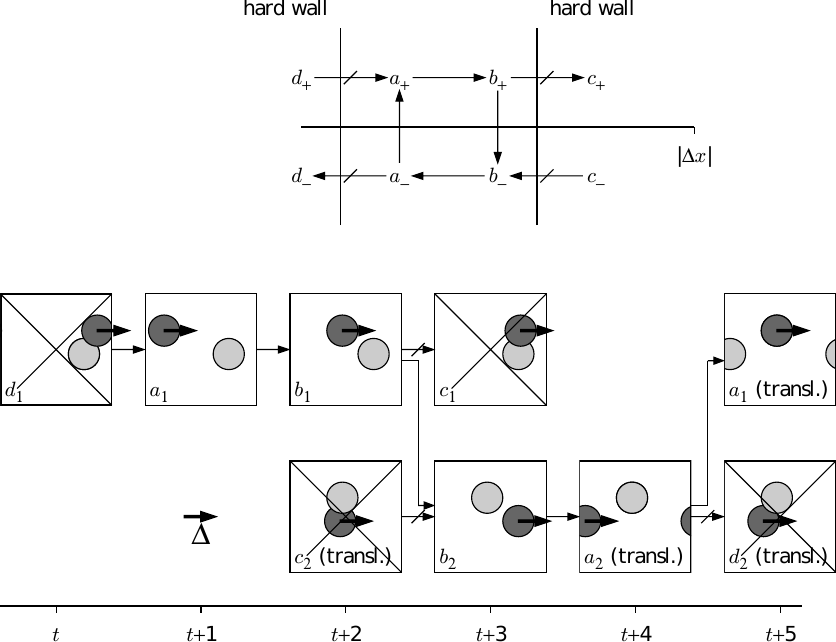} {
\emph{Upper}: Discrete one-dimensional system ($L=2$) with constant
probabilities $\pi(a)=\pi(b)$ and hard-wall conditions $\pi(c) =
\pi(d)=0$. The lifting variable $h = \pm 1$
corresponds to the direction of motion of the particle. \emph{Lower}: Equivalent
lifting algorithm for two
hard spheres with finite displacement $\Deltavec$ 
(the values of the lifting variable $\pm 1$ are replaced by $\{1,2\}$). The
forbidden particle moves $\TO{b_1}{c_1}$ and $\TO{a_2}{d_2}$ trigger lifting
moves. Maximal global balance is satisfied by moving in the $+x$ direction only.
The equivalence between one-dimensional motion and two-particle dynamics in a
constrained direction carries over to arbitrary pair potentials. 
}

As demonstrated in \fig{lifting_two_disks_movie}, the discrete one-dimensional system
with hard walls corresponds to two $D$-dimensional hard spheres that are
constrained to move only, say, along the $x$ direction, in a box with periodic
boundary conditions. The new lifting variable $i$ now indicates the moving sphere, and the
hard wall turns into a no-overlap condition for the two spheres.
Although the spheres only move towards the right, the algorithm satisfies
maximal global balance due to flows across the periodic boundaries.
Ergodicity for the unconstrained two-sphere problem in a $D$-dimensional box
is achieved by resampling, after a fixed number of steps, the moving particle
$i\in\{1,2\}$ and the direction of motion $\Deltavec \in \{+\xunitvec\Delta, +\yunitvec\Delta\}$ (for the
example of two hard disks in a periodic box).  The sequence of moves
between resampling is referred to in the following as an \emph{event
chain}.

\section{Infinitesimal moves and event-chain algorithm for interacting manyparticle systems}
\slabel{Nparticle}

We now extend the discussion of \sect{SingleParticle} to $N$-particle systems,
first for hard spheres, and then for particles with arbitrary pairwise potentials.
The idea to indicate the moving particle and its `momentum' by lifting variables
generalizes trivially to the $N$-particle case.  Special care is, however, required
to preserve global balance, and we show that the factorized Metropolis filter can
be used to implement maximal global balance in the infinitesimal-move limit.
We then implement this scheme efficiently in an event-driven MCMC algorithm.

Lifted configurations are now specified by the $N$ hard-sphere centers $(\rvec_1, \dots,
\rvec_N)$, the moving sphere $i$ and its direction of motion $\Deltavec$.  For concreteness,
we focus on the positive $x$ direction, $\Deltavec=+\xunitvec\Delta$, as before. A particle move is
\begin{equation}
a_i = (\rvec_1, \dots,\rvec_i, \dots, \rvec_N) \to 
b_i = (\rvec_1, \dots,\rvec_i + \Deltavec, \dots, \rvec_N).
\elabel{event_chain_move}
\end{equation}
This algorithm violates global balance because it generates
configurations with multiple overlaps, see
\fig{lifting_three_disks_movie}. In the presence of a multiple
overlap, it is impossible to define flows that satisfy the global
balance condition.  Multiple overlaps vanish, and maximal global
balance is recovered, for infinitesimal moves $|\Deltavec| \to 0$: In
that limit, the factorized Metropolis filter identifies a unique
collision partner, with probability one, since no two particles are at
the same distance from the moving particle.  The collision partner
then inherits the lifting variable and moves forward in the next
step. By a succession of infinitesimal steps that add up to a finite
chain displacement $\ell$, this reproduces the hard-sphere event-chain
algorithm\cite{Bernard_2009}.  Of course, the infinitesimal-move
algorithm is not implemented naively through a fine discretization,
but rather by identifying the next lifting event, and then advancing
the moving disk to contact.

\cfigure[50mm]{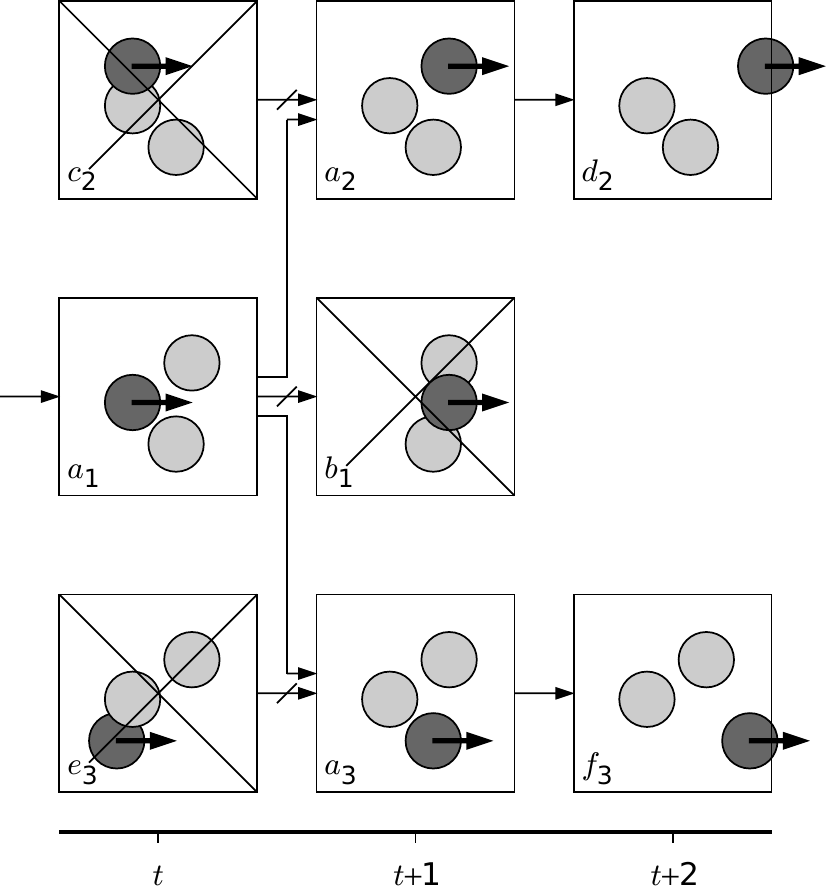}{Multiple overlaps for hard disks
(weight $\pi = \text{const}$ for the non-overlapping physical configurations 
$a,d,f$).
The violation of the global balance condition is
caused by the multiple overlap in configuration $b$ (overlap of
disk $1$ with both disks $2$ and $3$): the flow into all legal
configurations $a$, $d$, and $f$ must be equal, while the illegal (crossed-out)
configurations $b$, $c$, and $e$ generate zero flow. The multiple overlap
disappears, and global balance is again satisfied, in the limit $|\Deltavec| \to
0$. }

We now generalize the infinitesimal-move event-chain algorithm to arbitrary pair
potentials, using the factorized Metropolis filter,
\eq{pair_metropolis_filter}. For general interactions, the energy change between
configurations $a_i$ and $b_i$ which differ by an infinitesimal displacement
$\diff x_i$ of particle $i$ is
\begin{equation}
\diff E  = E(b) - E(a) = \sum_{j ( \ne i)} \frac{\partial E_{ij}(\vec r_j-\vec r_i)}{\partial x_i} \diff x_i = \sum_{j ( \ne i)} \diff E_{ij},
\end{equation}
where $\diff E_{ij}$ is the pairwise energy change, and $E_{ij}$ the pair
potential.
According to the factorized Metropolis filter, the move is rejected with probability
\begin{equation}
1-p^{\text{fact}}(\TO{a_i}{b_i}) =  1-\exp \Bigl(- \sum_{ j (\ne i) } \maxzero{
\diff E_{ij}} \Bigr) = \sum_{j (\ne i)} \maxzero{ \diff E_{ij}}.
\elabel{rejection_lifting}          
\end{equation}
Remarkably, for infinitesimal displacements, the rejection probability
is a sum of pair terms, while the individual
terms $\maxzero{\diff E_{ij}}$ normally neither add up to the total energy change $\diff E$ nor
to $\maxzero{\diff E}$. We use the terms in \eq{rejection_lifting} as the
probabilities for lifting moves
\begin{equation}
p^\text{lift}(\TO{a_i}{a_j}) =  \maxzero{\diff E_{ij}}\quad \forall j \ne i,
\elabel{lifting_infinitesimal}
\end{equation}
and obtain a rejection-free, maximal global balance MCMC algorithm.
\fig{lifting_three_particles} illustrates that the total flow into the
configuration $a_i$ equals the total flow out of configuration $a_i$,
satisfying the global balance condition \eq{GlobalBalance}.
Explicitly, the lifting flows in the example of \fig{lifting_three_particles} are:
\begin{align}
       \varphi_{a_2\to a_1} & = \pi(a)  \maxzero{ \diff E_{21}}\notag
\\
       \varphi_{a_3 \to a_1} & = \pi(a)  \maxzero{ \diff E_{31}}&
\notag
\intertext{and the particle move flow,}
       \varphi_{b_1 \to a_1} & =  
\varphi^\text{fact}_{a \to b} =  \pi(a)
(1 - \maxzero{  \diff E_{21}} -
\maxzero{ \diff E_{31}} ) \notag
\end{align}
Indeed, Eqs.~\eqref{e:rejection_lifting} and \eqref{e:lifting_infinitesimal}
define a rejection-free \emph{infinitesimal} MCMC algorithm
with maximal global balance.

\cfigure[45mm]{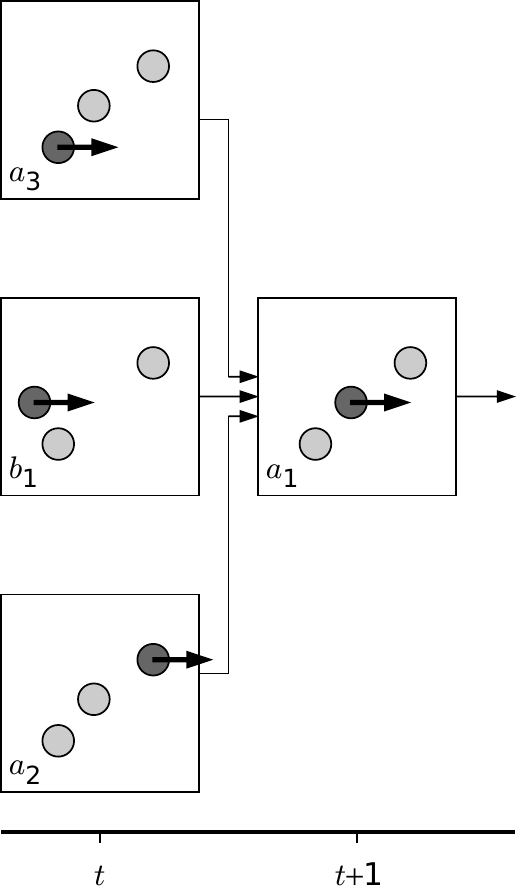}{Maximal global balance
for $N$ particles with arbitrary pair interactions (infinitesimal
step, factorized Metropolis filter of \eq{pair_metropolis_filter}). 
Flow into $a_1$ is due to $N-1$ lifting moves (here, for $N = 3$,
$\TO{a_2}{a_1}$ and
$\TO{a_3}{a_1}$) and to a particle move (here, $\TO{b_1}{a_1}$). 
For infinitesimal steps, the flow into $a_1$
equals $\pi(a)$ (see text), thus balancing the flow $\pi(a)$
out of $a_1$ and satisfying global balance, \eq{GlobalBalance}.}

In order to implement this algorithm efficiently, we choose an
event-based approach. As mentioned, the factorized Metropolis filter ensures that no
two lifting events can occur in the same infinitesimal timestep.
Thus, every interaction of the moving particle with another can be
treated independently of other interactions.
Further, for fixed partner $j$, the lifting
probabilites at successive timesteps are independent, and they vanish
if the pair potential decreases during displacement. Following the
BKL algorithm\cite{Bortz_1975,Peters_2012}, we
determine the displacement until the first lifting move occurs by
sampling a uniform random number $\Upsilon_{ij}$ from $(0,1]$ which
determines the admissible energy increase until lifting,
$E^*_{ij}=-\ln\Upsilon_{ij}$.  The displacement until lifting $s_{ij}$
is then found from
\begin{equation}
    E^*_{ij}=\int_0^{E^*_{ij}} \maxzero{\diff E_{ij}}
    = \int_0^{s_{ij}} \left[\frac{\partial E_{ij}(\vec r_j-\vec r_i-s\xunitvec)}{\partial s}\right]^+ \diff s.
\label{eqFindCollision}
\end{equation}
If this equation lacks a solution due to the shape of the interaction
potential, or due to a large thermal excitation $E_{ij}^*$,
no lifting event is generated, $s_{ij}=\infty$.  While solving
Eq.~\eqref{eqFindCollision} can be nontrivial in general, we give a
fast method for the most usual pair potentials below.  The smallest
of the $N-1$ independent $\{s_{ij}\}_{j \neq i}$ determines the lifting move $i\to
j^*$ which will actually take place, advancing the moving disk by
$\min_{j\neq i}(s_{ij})=s_{ij^*}$ in the $+\xunitvec$ direction and changing the lifting
variable to $j^*$.  We also make sure that the total displacement in a
single event chain equals the chain displacement $\ell$, which usually
requires truncating the final event.  After the end of the chain, the
lifting variables are resampled.  Each chain thus consists in an
infinite succession of infinitesimal moves that add up to the chain
displacement $\ell$. Alternatively, we could introduce a small
constant probability for terminating a chain in each infinitesimal
move. This would effectively lead to exponentially distributed random
$\ell$, and is also a valid MCMC algorithm.

We conclude with some practical remarks on solving
Eq.~\eqref{eqFindCollision} for model potentials that occur in
practice.  Many model potentials are central potentials, $E_{ij}(\vec
r_{ij}) = E_{ij}(r_{ij})$.  If the pair potential consists of several
terms, e.~g.~attractive and repulsive terms, it may be convenient to
treat them separately by a further factorization of the Metropolis
filter, and decompose the lifting probabilities, $\maxzero{\diff
  E^{\text{attr}}_{ij}} + \maxzero{\diff E^{\text{rep}}_{ij}}$, where
$E^{\text{attr}}$ and $E^{\text{rep}}$ are the attractive and
repulsive parts of the pair potential. This decomposition can lead,
however, to a higher event rate than the full potential. For instance, the
Lennard-Jones potential reduces to two soft-sphere interactions,
one attractive and one repulsive, and the mean free path between events is
reduced at most by half in comparison to the case without
decomposition.  In return, the decomposition greatly simplifies the
Monte Carlo program.

We will thus focus here on the case of soft-sphere potentials
which are monotonous.  A lifting move can only be generated if the moving
particle is in the rising part of the pair potential.
In this case, solving Eq.~\eqref{eqFindCollision} amounts
to sampling the energy increase $E^*_{ij} = -\log \Upsilon_{ij}$, with $\Upsilon_{ij}$
a uniform random number from $(0,1]$, which fixes the interaction energy
$E^{\text{lift}}_{ij} = E_{ij}(r_{ij}) + E^*_{ij}$, and thus the interparticle distance 
$r_{ij}^{\text{lift}}=E^{-1}_{ij}(E^{\text{lift}}_{ij})$ at the lifting move.
(If $E^{\text{lift}}$ exceeds any possible value of the interaction potential,
there is no lifting move generated.)  The admissible displacement $s_{ij}$
for the $i,j$ particle pair is then the positive root of 
$r_{ij}^{\text{lift}}=|\vec r_j-\vec r_i-s_{ij}\xunitvec|$.  Again, if no such
root exists, no lifting move is generated, and particle $i$ will pass
particle $j$.  Using this method, and the decomposition into attractive
and repulsive terms, Eq.~\eqref{eqFindCollision} is thus easily computable for
a large range of potentials.

In systems with periodic boundary conditions, for very long chains, a
particle can pass by the same collision partner more than once: The
pair potential no longer is monotonous.  This is most easily avoided
by tuning the chain displacement so that the moving sphere can only
interact with one periodic image of the other spheres or by
introducing a lifting move of the moving sphere with itself after a
displacement of half the box. This move does not change the statistics
of the following events. After it, the next event is computed as usual.

\section{Speedup with respect to Metropolis Monte Carlo}

We now compare the performance of the generalized ECMC algorithm with
Metropolis Monte Carlo (MMC).  As an application, we consider a
two-dimensional system of $N$ particles interacting with a truncated
pairwise power-law potential, $E_{ij}(r) = \tilde E(\operatorname{min}
(r, r_{\rm c}))$, with $\tilde E(r) = \epsilon (\sigma/r)^n$ and
$r_{\rm c} = 1.8 \sigma$, $\sigma$ being the particle diameter. This
potential includes important physical interactions such as the dipole
interaction in magnetic colloids\cite{MagneticCollids}, hard disks
($n\rightarrow\infty$) and Lennard-Jones particles, once decomposed
into a repulsive soft-sphere interaction ($n=12$) and an attractive
one ($n=6$).  In comparison to MMC, the ECMC algorithm uses more
random numbers, one per interaction term, whereas Metropolis uses one
per step.  In our implementations, however, the main computational
workload is the evaluation of the potentials, not the generation of
random numbers (using the Mersenne Twister).  One event of the ECMC
algorithm is thus implemented in the same amount of time as one
attempted step of MMC.  ($3.2\cdot 10^9$ steps per hour in MMC,
$1.5\cdot 10^9$ events per hour in ECMC). We compare the performance
of the algorithms in terms of the CPU time used (see \fig{speedup2}
for details).

As estimate of the relative performance of the algorithms
we consider as in other recent work\cite{Bernard_2009,Kapfer_2013,Engel_2013}
the autocorrelation time $\tau_6$ of the global orientational order parameter $\Psi_6$,
\begin{equation}
    \Psi_6 = \frac{1}{N} \cdot \sum_{i,j} \frac{A_{ij}}{A_i} \exp(6\mathrm i \theta_{ij})
\end{equation}
where $\theta_{ij}$ is the angle of the bond vector between particles
$i$ and $j$ against a fixed axis and $A_{ij}/A_i$
the contribution of particle $j$ to particle $i$'s Voronoi cell
perimeter \cite{Strandburg,Mickel_2013}.  With $\Psi_6$ being a global observable,
we assume that its autocorrelation time $\tau_6$ is representative for the mixing time
for dense liquid states close to the freezing point,
located at $\rho\sigma^2=1$ for $n=12$, and $\rho\sigma^2\approx 0.89$ for harder
disks with $n=1024$, where $\rho\sigma^2$ is the dimensionless density, $N\sigma^2/V$.

We find that in terms of CPU time, in small systems, ECMC mixes a few
times quicker than MMC.  We tuned both algorithms to their optimal
parameters (see \fig{speedup2} for details).  Speedup, defined as the
ratio $\tau_6(\text{MMC})/\tau_6(\text{ECMC})$, is, in the region of
study, not found to be a strong function of density (\fig{speedup2},
bottom row).  For increasing system size, however, the speedup increases
(\fig{speedup2}, top row).  An increase of speedup with system size
has also been found for hard-sphere systems, where it approaches two
orders of magnitude in large systems of $10^6$ particles
\cite{Engel_2013}.  We thus expect that the generalized ECMC algorithm
has similar characteristics with respect to system size as the
hard-sphere ECMC algorithm.  For very large systems, the MMC algorithm
does not equilibrate within the allotted simulation time: The
distribution of $|\Psi_6|$ is not yet stationary, even though the
simulation time exceeds the $\tau_6$ by a factor of 100.  Thus, we
have not determined $\tau_6$ for these systems.

\cfigure[85mm]{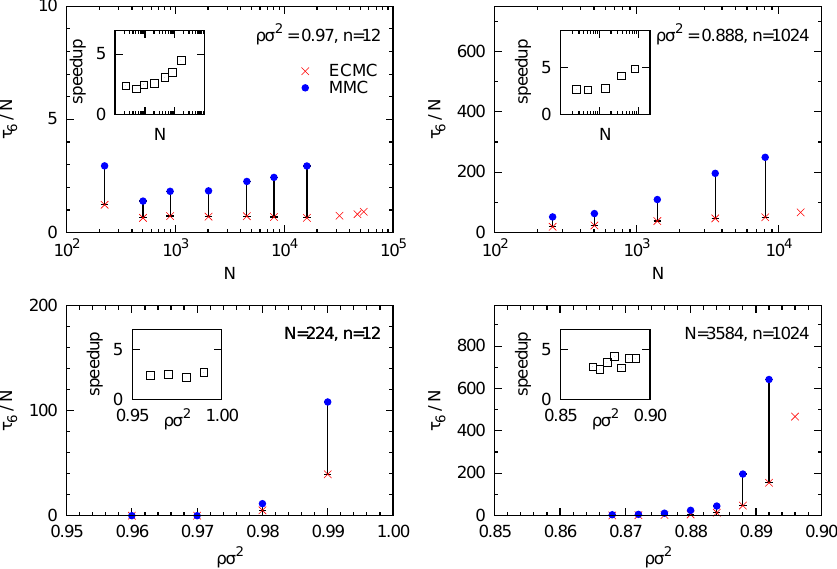}
{Autocorrelation times $\tau_6$ of the $\Psi_6$ orientational order parameter
for event-chain Monte Carlo (ECMC) and
Metropolis Monte Carlo (MMC), for soft-disk systems of $N$ particles at inverse
temperature $\beta=1$.
$\tau_6$ is measured in arbitrary though comparable amounts of CPU time:
One unit of CPU time
for ECMC is a displacement of $T=100N\sigma$, chain displacement
$\ell = 0.025\sqrt{N}\sigma$, spanning thus about half the system volume.
Event chains in $+x$ and $+y$ alternate every $\Delta T = N\sigma /2$ of displacement.
One unit of CPU time for Metropolis consists of $1000N$ moves, where a move
is the attempt to displace a particle by a random vector sampled from a
disk of radius $0.16 \sigma$.}

\section{Direct pressure computation}
\slabel{secPressure}

In order to obtain the equation of state in the $NVT$ ensemble for the particle system under study,
the pressure $P$ must be computed. Usually, the pressure is
obtained using the virial theorem (see Sec.~2.2 of \cite{HansenMcdonald}),
either by averaging the virial, or by integration of the product of
the static pair correlation function $g(r)$ and the pair potential $E_{ij}(r)$.
Direct averaging is not possible for hard-sphere interactions, since the
potential is singular. It is thus required to compute a discrete approximation
of $g(r)$ and extrapolate it to the contact value to obtain $P$.  Even for
non-singular steep potentials, the approach via $g(r)$ is bothersome, since
the dominant contributions to $P$ come from close pairs (for the family of power-law
potentials, $r\approx\sigma$), which is poorly sampled in the canonical ensemble.
Finally, evaluation of the virial during the simulation implies extra computation
for evaluating the forces.  By contrast, in hard-sphere event-driven molecular dynamics
the virial pressure is directly related to the collision rate, which is a trivial
byproduct of the computation \cite{AlderWainwrightMD}:
\begin{equation}
   \beta P = \rho - \frac{\beta \rho m}{2 \tsim N} \sum_{\text{collisions}} b_{ij},
   \elabel{event_chain_md_pressure}
\end{equation}
where $\rho=N/V$ is the particle number density, $\tsim$ the total simulation time, 
$m$ the mass of a particle, and $b_{ij} = (\vec
r_i-\vec r_j)(\vec v_i-\vec v_j)$, with $\vec r_{i,j}, \vec v_{i,j}$ the positions
and velocities of the colliding particles.  In the following we show that in ECMC,
the rate of lifting moves is, just like the collision rate in event-driven MD,
directly related to the pressure.  We give an elementary derivation
independent of the virial theorem for the soft-particle case.  The results are,
however, also valid for hard particles and can be derived using arguments
by Speedy \cite{Speedy_1988} which connect the pressure to the stochastic geometry
of the admissible configurations.

\cfigure[50mm]{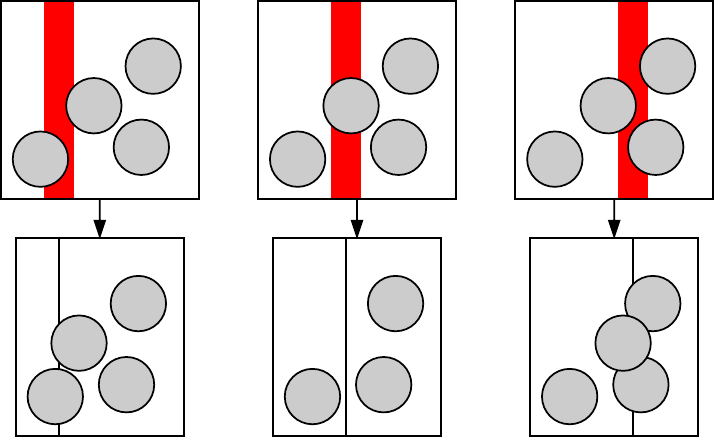}{Virtual rift volume changes by random
removal of an infinitesimal strip from a hard-sphere configuration.
\emph{Left}: A successful removal, \emph{Center}: Elimination of a
particle (ideal gas pressure), \emph{Right}: Generation of an overlap
(excess pressure). The left and right cases become indistinct for 
soft interactions.}  

\cfigure[50mm]{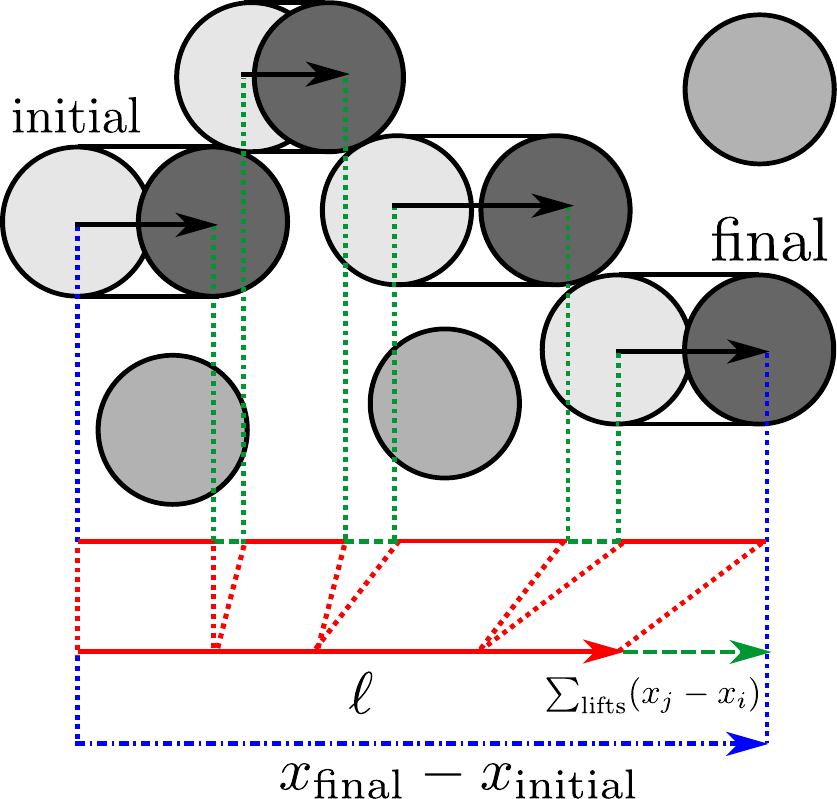}{Direct computation of the pressure: The excess pressure
  is derived from the ratio of excess displacement (green dashed lines,
  $\sum_{\text{lifts}} (x_j-x_i)$)  and the chain displacement $\ell$ (red solid line).
  For isotropic systems, only the distance between the final and initial
  particle $x_{\text{final}}-x_{\text{initial}}$
  (blue dash-dot arrow) has to be recorded.}

In order to compute the pressure $\beta P = \partial \ln Z/\partial V$,
we consider virtual \emph{rift volume changes} effected by removing
a randomly located strip of size $\diff L_x \times L_y$ from the system
(see \fig{xpressure_virtual_move}).  By considering all positions of
the strip, this procedure yields all $N$-particle configurations in the smaller
simulation box, and thus the new partition function $Z(V+\diff V)$. For isotropic
systems, we recover the virial expression,
\begin{equation}
    \beta P = \rho + \frac 1V \left\langle \sum_{\{i,j\}} (x_j-x_i)
    \frac{\beta\partial E_{ij}(\vec r_i - \vec r_j)}{\partial x_i}\right\rangle,
\label{eqIsoPressure}
\end{equation}
where $\langle\cdot\rangle$ is the canonical average.
The first term of the rhs in Eq.~\eqref{eqIsoPressure} is due to particles located
in the removed strip, which lead to illegal configurations with
less than $N$ particles (\fig{xpressure_virtual_move}, center). This
term yields the ideal-gas pressure.  The non-ideal contribution to the
pressure results from changes in the Boltzmann weight due to
compressed bonds, with $(x_j-x_i)$ accounting for the probability of a
bond traversing the removed strip. In hard spheres, this term is
produced by particle overlap (\fig{xpressure_virtual_move}, right
panel). Replacing the canonical average in Eq.~\eqref{eqIsoPressure}
by an average in the lifted canonical ensemble, $\langle
\cdot\rangle_k := N^{-1} \sum_k \langle \cdot \rangle $, where $k$ is
the lifting variable, one of the sums collapses and yields a factor of
$N$; we recover the probabilities for a lifting move from $i\to j$ and
$j\to i$. An ECMC simulation will reproduce the lifted canonical
average and thus yields an unbiased estimator of the pressure.
Summing up all the lifting events in a chain (see \fig{excess_displacement}), we obtain
\begin{equation}
    \beta P =  \rho  \cdot \left\langle
         \frac{x_{\text{final}} - x_{\text{initial}}}{\ell} \right\rangle_{\text{chains}},
\elabel{ecmcPressure}
\end{equation}
where $\left\langle\cdot\right\rangle_{\text{chains}}$ is the
average over event chains, $x_\text{initial}$ is the position of
the first particle before the effects of the chain, and
$x_\text{final}$ the position of the last particle after, adjusted for
periodic boundaries if necessary.  Thus, it suffices to know the
beginning and end of event chains to compute the
pressure. Explicitly,
\begin{equation}
  x_{\text{final}} - x_{\text{initial}} = \ell + \sum_{\text{lifts}} (x_j-x_i)
\elabel{ecmcexcess}
\end{equation}
where $x_i$ and $x_j$ are the positions of the moving particle $i$ and
of the hit particle $j$, respectively, \emph{at lifting}, see \fig{excess_displacement}.
In the ideal gas, there are no lifting moves and Eq.~\eqref{e:ecmcPressure} reduces to
the ideal gas pressure. The excess displacement $(x_j-x_i)$ can be
negative for an interaction potential with attractive components, such as
Lennard-Jones.  If the potential is decomposed into attractive and repulsive
parts as outlined in \sect{Nparticle}, individual excess displacements for the 
two potentials also add up to the correct pressure.  As
evidenced by \tab{pressure_comparison}, the results obtained from ECMC
via Eq.~\eqref{e:ecmcPressure} agree with the
conventional virial approach.  Since no extra computation is required,
the procedure via the excess displacement in ECMC is more efficient
than the virial approach, in particular for steep potentials.

Finally, one might be interested in anisotropic systems where the
collision rates can depend on the direction of the event
chains.  In this case, the derivation presented for
longitudinal rifts (removing strips normal to the chain direction) is
supplemented with an analogous result for transverse rifts (removing
strips aligned with the event chain), which leads in $D$ dimensions to
the full pressure,
\begin{align}
    \beta P =  \rho  +  \left\langle
         \frac{\rho}{D\ell} \sum_{\text{lifts}}\frac{(\vec r_{j}-\vec r_{i})^2}{x_j-x_i}
     \right\rangle_{\text{chains}},
\end{align}
where $x$ is the coordinate parallel to the chain direction.
More generally, the full stress tensor $\mathbf \tau$ can be computed as an average of
the dyadic product of the interparticle distance $\vec r_{ij}=\vec r_j-\vec r_i$ at collision:
\begin{align}
    \beta \mathbf \tau &= -\rho \mathbf 1
    -
    \left\langle
     \frac{\rho}{\ell} \sum_{\text{lifts}}\frac{
            \vec r_{ij}^{} \vec r_{ij}^{\rm t} }{x_j-x_i}
     \right\rangle_{\text{chains}},
\end{align}
where $\mathbf 1$ is the identity matrix in $D$ dimensions.

\begin{table}
\begin{tabular}{lcccc}
  $n$      & $\quad N \quad $     & $\rho\sigma^2$  & virial pressure        & ECMC pressure \\ \hline
 $ 12 $    & $2^{14}$ & $0.990$      & $14.4369 \pm 0.0058 $  & $14.4267 \pm 0.0038$ \\
 $ 48 $    & $2^{14}$ & $0.860$      & $8.753 \pm 0.011$      & $8.7565 \pm 0.0023$ \\
 $48  $    & $2^{14}$ & $0.888$     & $9.441 \pm 0.025$      & $9.429 \pm 0.027$ \\
 $1024$    & $2^{14}$ & $0.888$     & $9.174 \pm 0.028$      & $9.1679 \pm 0.0026$ \\ 
  $\infty$ (HS) & $2^{16}$ & $0.888$  & $9.1667 \pm 0.0073$  & $9.1723 \pm 0.0064$ \\\hline
12, 6 (LJ)  & $2^{14}$ & $0.888$     & $1.44833 \pm 0.00031$ & $1.447623 \pm 0.000045$
\end{tabular}

\caption{\label{ecmcPressureVerif}Comparison of pressure computed using
the virial expression and from excess displacement in ECMC \eq{ecmcPressure},
for repulsive soft and hard sphere (HS) interactions, and for the
Lennard-Jones (LJ) potential, at $\beta=1$.  Pressures and densities are
nondimensionalized,
$\beta P \sigma^2$ and $\rho\sigma^2$. 
The deviations given are standard errors from 10 independent simulations each.
For LJ, the potential was decomposed into attractive and repulsive parts.
\tlabel{pressure_comparison}}
\end{table}

\section*{Conclusion}

In the present article, we have generalized the event-chain Monte Carlo
algorithm from hard spheres to particle systems interacting with arbitrary pair
potentials, such as Lennard-Jones liquids or soft disks. 
The resulting algorithm is faster than conventional Metropolis Monte Carlo, 
with the gap in performance increasing with the system size.  It is based
on the lifting concept, and relies on a new factorization of the Metropolis
filter, applied to infinitesimal Monte Carlo moves, to achieve maximal global balance.
The infinitesimal moves are implemented efficiently in an event-based algorithm
using the BKL approach.  The algorithm generates a continuum of samples of the
equilibrium distribution. This has allowed us to derive the pressure and the stress
tensor in the $NVT$ ensemble directly from the simulation without any
additional computation.  Even though presented in periodic boundary conditions,
the algorithm also applies to nonperiodic systems, by introducing chains in the
$-x$ and $-y$ direction to render it ergodic.

Infinitesimal moves permit to apply the framework of lifted Markov
chains to the interacting particles problem, since they define
uniquely the next event, while satisfying global balance.  By
subdivision into infinitesimal moves, both the original hard-sphere
event-chain algorithm\cite{Bernard_2009}, and the hybrid MC algorithm
of Peters and de With\cite{Peters_2012} are revealed to be lifting
algorithms.  Lifting improves mixing in large, strongly correlated
systems, since clusters of particles are displaced in a cooperative
way.  It is thus applicable to packing problems, glassy systems, etc.
particularly, as its dynamics are fundamentally different from the MMC
or MD case.  As the original event-chain algorithm, it can be parallelized
\cite{Kapfer_2013}.  We expect the algorithm to extend to complex
fluids, with particles possessing internal degrees of freedom, to path
integral (quantum) Monte Carlo, and other sampling
problems.

\begin{acknowledgments}
The authors thank E.~A.~J.~F.~Peters and M.~Engel for
useful discussion and comments on a previous version of the manuscript.
\end{acknowledgments}

\end{document}